\documentstyle[aps,prl,preprint,floats,epsf]{revtex}

\begin{document}

%\begin{flushright}
%SLAC-PUB-8047\\
%CSUHEP 99/01\\
%BaBar Note \#481\\
%hep-ph/9902313\\
%February 1999
%\end{flushright}

%\tighten

\title{Discrete Ambiguities in the Measurement of the Weak Phase $\gamma$}

\author{Abner Soffer\footnote{Permanent address: Department of Physics, 
	Colorado State University, Fort Collins, CO 80523.\\
	Work supported by Department of Energy contracts DE-AC03-76SF00515 
	and DE-FG03-93ER40788, and by the National Science
Foundation }}

\address{Stanford Linear Accelerator Center, Stanford University, Stanford,
	 CA 94309, USA}
\date{\today}
\maketitle

\begin{abstract}
Several time-independent methods have been devised for measuring the
phase $\gamma$ of the Cabibbo-Kobayashi-Maskawa~\cite{ref:ckm}
unitarity triangle. It is shown that such measurements generally
suffer from discrete ambiguity which is at least 8-fold, not 4-fold as
commonly stated. This has serious experimental implications, which are
explored in methods involving $B\rightarrow DK$ decays. The
measurement sensitivity and new physics discovery potential are
estimated using a full Monte Carlo detector simulation with realistic
background estimates.
\end{abstract}


\section{Discrete Ambiguities}	
Direct CP-violation within the standard model takes place when two or
more amplitudes with different CKM phases interfere in a
time-independent manner. As an example,
take the decay of a $B^+$ meson to a final state $f^+$, which can
proceed through two amplitudes. The partial decay widths can be
written as
%%%%%%%%%%
\begin{equation}
\Gamma(B^\pm\rightarrow f^\pm) = A_1^2 + A_2^2 + 
	2 A_1 A_2 \cos(\delta \pm \phi), \label{eq:interf}
\end{equation}
%%%%%%%%%%%%%%%%%%%
where $A_1$ and $A_2$ are the magnitudes of the interfering
amplitudes, $\phi$ is the CKM phase difference between the amplitudes,
and $\delta$ is the CP-conserving phase difference. CP-violation is
often considered in terms of a non-vanishing CP-asymmetry,
%%%%%%%
\begin{equation}
{\cal A} = {\Gamma(B^+\rightarrow f^+) - \Gamma(B^-\rightarrow f^-) \over
	    \Gamma(B^+\rightarrow f^+) + \Gamma(B^-\rightarrow f^-)}
         = -{2 R \sin\delta \sin\phi \over
	    1 + R^2 + 2 R \cos\delta \cos\phi},\label{eq:asym}
\end{equation}
%%%%%%%%%%%%%%%%%%%
where $R = A_2/A_1$. It is clear that ${\cal A}$ gives a determination
of $\gamma$ which has a 4-fold ambiguity,
due to its invariance under the two symmetry operations
%%%%%%%%%%%%%
\begin{eqnarray}
S_{\rm sign} &:&\gamma \rightarrow -\gamma,\ \ \delta_B \rightarrow -\delta_B 
	\nonumber \\
S_{\rm exchange} &:& \gamma \leftrightarrow \delta_B.
	\label{eq:ambig}
\end{eqnarray}
%%%%%%%%%%%%%%%%%%%%%%%%
These ambiguities were also noted in methods which do not rely on 
a decay rate asymmetry~\cite{ref:GW}. 

The $S_{\rm sign}$ ambiguity is generally considered unphysical, since
application of $S_{\rm sign}$ to $\gamma$ values within the currently
allowed range~\cite{ref:buras},
%%%%%
\begin{equation}
g \equiv \{40^\circ \lesssim \gamma \lesssim 100^\circ\}, 
\end{equation}
%%%%%%%%%%%%%%
yields values which are not within $g$, and
are therefore inconsistent with the standard model. 
We note, however, a third symmetry of Equation~(\ref{eq:interf}, which 
makes the ambiguity 8-fold:
%%%%%
\begin{equation}
S_\pi : \gamma \rightarrow \gamma + \pi, \ \ 
	\delta_B \rightarrow \delta_B - \pi.
	\label{eq:ambig2}
\end{equation}
%%%%%%%%%%%%%
This symmetry has so far been overlooked, perhaps because it does
not appear to be a symmetry of ${\cal A}$

The significance of $S_\pi$ arises from the fact that while $S_\pi$
and $S_{\rm sign}$ may be unphysical within the standard model, the
operation $S_\pi S_{\rm sign}$, applied to $\gamma \in g$, results in
values which are in or close to $g$. Thus, given the finite
sensitivity of future experiments, it may be impossible to deem $S_\pi
S_{\rm sign}$ unphysical, hampering the ability to test the standard
model using such measurements of $\gamma$.

In some $\gamma$ measurement methods, there is no a-priori knowledge
of the magnitudes of some of the interfering amplitudes. These
magnitudes are then obtained simultaneously wth $\gamma$. In this
case, additional ``accidental'' ambiguities will exist if more than one
magnitude value is consistent with the data. An example is given
below.

\section{Quantitative Experimental Implications}
$B\rightarrow DK$ decays, recently observed by CLEO~\cite{ref:b2dk},
provide several ways to measure the weak phase $\gamma = \arg(-V_{ud}
V_{ub}^* / V_{cd} V_{cb}^*)$. Since non-standard model effects are
expected to be small in such decays, comparing these measurements with
experiments which are more sensitive to new physics may be used to
test the standard model~\cite{ref:grossman}. Gronau and Wyler
(GW)~\cite{ref:GW} have proposed to measure $\gamma$ in the
interference between the $\bar b\rightarrow \bar c u\bar s$ decay
$B^+\rightarrow \bar D^0 K^+$ and the color-suppressed, $\bar
b\rightarrow \bar u c \bar s$ decay $B^+\rightarrow D^0 K^+$.
Interference occurs when the $D$ is observed as one of the
CP-eigenstates $ D^0_{1,2} \equiv {1\over\sqrt{2}} \left(D^0 \pm \bar
D^0\right)$, which are identified by their decay products. The
interference amplitude is
%%%%%
\begin{equation}
\sqrt{2}\; A(B^+\rightarrow \bar D^0_{1,2} K^+) = 
  \sqrt{{\cal B}(B^+\rightarrow D^0 K^+)} e^{i(\delta_B + \gamma)} \pm 
  \sqrt{{\cal B}(B^+\rightarrow \bar D^0 K^+)},
\label{eq:GW-triangle}
\end{equation}
%%%%%%%%%%%%%%%%%%%%%%%
where $\delta_B$ is a CP-conserving phase. The value of $\gamma$ is
extracted from this triangle relation and its CP-conjugate,
disregarding direct CP-violation in
$D^0$~decays~\cite{ref:grossman2}. Several variations of the method
have been developed\cite{ref:dunietz,ref:resonance}.

In practice, measuring the branching fraction ${\cal B}(B^+\rightarrow
D^0K^+)$ requires that the $D^0$ be identified in a hadronic final
state, $f=K^-\pi^+ (n\pi)^0$, since full reconstruction is impossible
in semileptonic decays, resulting in unacceptably high
background. Atwood, Dunietz and Soni (ADS)~\cite{ref:ads} pointed out
that the decay chain $B^+\rightarrow D^0 K^+$, $D^0\rightarrow f$
results in the same final state as $B^+\rightarrow \bar D^0 K^+$,
$\bar D^0\rightarrow f$, where the $\bar D^0$ undergoes doubly Cabibbo
suppressed decay. Estimating the ratio between the interfering decay
chains, one obtains
%%%%%%
\begin{eqnarray}
\left|{A(B^+\rightarrow \bar D^0 K^+) \; 
	A(\bar D^0\rightarrow f)
	\over
A(B^+\rightarrow D^0 K^+) \; A(D^0\rightarrow f)}\right|
 &\approx&
	\left|{V_{cb}^* \over V_{ub}^*}  \;
	{V_{us}  \over V_{cs}}\;
	{a_1 \over a_2}\right| \;
	\sqrt{{\cal B}(\bar D^0 \rightarrow f) \over 
		{\cal B}(D^0 \rightarrow f)} \approx 0.6.
\label{eq:GW-trouble}
\end{eqnarray}
%%%%%%%%%%%%%% 
The numerical value in Equation~(\ref{eq:GW-trouble}) was obtained using
$|{V_{cb}^* / V_{ub}^*}| = 1/0.08$~\cite{ref:cleo-Vub}, 
$|{V_{us}  /V_{cs}}| = 0.22$, 
$|a_1 /a_2| = 1/0.26$~\cite{ref:cleo-bsw}, and
%%%%%%%
\begin{equation}
{{\cal B}(\bar D^0 \rightarrow f) \over {\cal B}(D^0 \rightarrow f)}
= 0.0031, \label{eq:dcs}
\end{equation}
%%%%%%%%%%%%%%%%%
which is the ratio measured for $f = K^-\pi^+$~\cite{ref:gronberg}.
Equation~(\ref{eq:GW-trouble}) implies that sizable interference makes
it practically impossible to measure ${\cal B}(B^+\rightarrow
D^0K^+)$, and the GW~method fails.

ADS proposed to use the interference of Equation~(\ref{eq:GW-trouble})
to obtain $\gamma$ from the decay rate asymmetries in $B^+\rightarrow
f_iK^+$, where $f_i, \ i = 1,2,$ are two $D$ final states of the type
$K^-\pi^+ (n\pi)^0$. Measuring the four branching fractions, ${\cal
B}(B^+ \rightarrow f_i K^+)$, ${\cal B}(B^- \rightarrow \bar f_i
K^-)$, one calculates the four unknowns ${\cal B}(B^+\rightarrow D^0
K^+)$, $\gamma$, and the two CP-conserving phases associated with the
two decay modes. ${\cal B}(B^+\rightarrow \bar D^0 K^+)$ and the $D^0$
decay branching fractions will have already been measured to high
precision by the time the rare decays $B^+\rightarrow f K^+$ are
observed. In addition to the similar magnitudes of the interfering amplitudes,
large CP-conserving phases are known to occur in
$D$~decays~\cite{ref:ddecays}, making large decay rate asymmetries
possible in this method. 

Jang and Ko (JK)~\cite{ref:jang-ko} and Gronau and Rosner~\cite{ref:gr}
have developed a $\gamma$ measurement method similar to the GW method,
but in which ${\cal B}(B^+\rightarrow D^0 K^+)$ is not measured
directly. Rather, it is essentially inferred by using the larger branching
fractions of the decays $B^0\rightarrow D^- K^+$, $B^0\rightarrow \bar
D^0 K^0$ and $B^0\rightarrow D_{1,2} K^0$, solving in principle the
problem presented by Equation~(\ref{eq:GW-trouble}). 

\section{Combining the ADS and the GW Methods}
Since the $\bar b\rightarrow \bar u c \bar s$ amplitude in
$B\rightarrow DK$ is very small and hard to detect, several methods
will have to be combined in order to make best use of the limited
data. Quantitative estimates of the resulting gain in sensitivity are
rarely conducted, since they require realistic efficiency and
background estimates, and depend on specific phase values. Here
we undertake this task for the case of combining the ADS and GW
methods (contributions of the JK method are commented on
later). In this scheme, one obtains the unknown parameters
%%%%%%%%%%%%
\begin{equation}
\xi \equiv\left\{
{\cal B}(B^+\rightarrow D^0 K^+), \ \gamma,\ \delta_B, \ \delta_D\right\},
\end{equation}
%%%%%%%%%%%%%%%%%%%
where $\delta_D = \arg[A(D^0\rightarrow f)A(\bar D^0\rightarrow f)^*]$,
by minimizing the function
%%%%%%%%%%%%
\begin{equation}
\chi^2(\xi) = \left({a(\xi) - a_m\over \Delta a_m}\right)^2+ 
	\left({\bar a(\xi) - \bar a_m\over \Delta \bar a_m}\right)^2 + 
	\left({b(\xi) - b_m\over \Delta b_m}\right)^2 + 
	\left({\bar b(\xi) - \bar b_m\over \Delta \bar b_m}\right)^2
	\label{eq:chi2}
\end{equation}
%%%%%%%%%%%%%%%%%%
with respect to the parameters $\xi$.
In Equation~(\ref{eq:chi2}) we use the symbols
%%%%%%
\begin{eqnarray}
a_m      &\equiv& {\cal B}(B^+\rightarrow f K^+) \nonumber\\
b_m      &\equiv& {\cal B}(B^+\rightarrow D^0_{1,2} K^+)
\end{eqnarray}
%%%%%%%%%%%%%%
to denote the experimentally measured decay rates of interest, and
%%%%%%%%
\begin{eqnarray}
a(\xi) &\equiv& \left|\sqrt{{\cal B}(B^+\rightarrow \bar D^0 K^+) \; 
	               {\cal B}(\bar D^0\rightarrow f)} +
	       \sqrt{{\cal B}(B^+\rightarrow      D^0 K^+) \; 
	             {\cal B}(D^0\rightarrow f)}
	 e^{i(\delta_D + \delta_B + \gamma)}
	\right|^2 \nonumber\\
b(\xi) &\equiv& {1 \over 2}
	\left| \pm \sqrt{{\cal B}(B^+\rightarrow \bar D^0 K^+)} +
	         \sqrt{{\cal B}(B^+\rightarrow      D^0 K^+)}
	 e^{i(\delta_B + \gamma)}
	\right|^2\label{eq:theo-quan}
\end{eqnarray}
%%%%%%%%%%%%%%%%
to denote the corresponding theoretical quantities. $\bar a_m$, $\bar b_m$,
$\bar a(\xi)$ and $\bar b(\xi)$ are the CP-conjugates of $a_m$, $b_m$,
$a(\xi)$ and $b(\xi)$, respectively. $\Delta x_m$ represents the
experimental error in the measurement of the quantity $x_m$.
%The
%branching fractions ${\cal B}(D^0\rightarrow f)$, ${\cal B}(\bar
%D^0\rightarrow f)$ and ${\cal B}(B^+\rightarrow \bar D^0 K^+)$ are
%taken from other, high-statistics measurements.

Several gains over the individual methods are immediately apparent: In
the ADS~method, a $D$ decay mode is ``wasted'' on measuring the
uninteresting CP-conserving phases. By contrast, when combining the
methods, knowledge of $a_m$, $b_m$, $\bar a_m$ and $\bar b_m$ in a
single mode is in principle enough to determine the four unknowns,
$\xi$, even if $\delta_D = \delta_B = 0$. In practice, adding the
$D^0_{1,2}$ modes will decrease the statistical error of the
measurement. In both the GW and the ADS methods, the ability to
resolve the $S_{\rm exchange}$ ambiguity depends on the degree to
which $\delta_B$ varies from one $B^+$ decay mode to the
other. Experimental limits on CP-conserving phases in $B\rightarrow
D\pi$, $D^*\pi$, $D\rho$ and $D^*\rho$~\cite{ref:fsi} suggest that
$\delta_B$ may be small, making the $S_{\rm exchange}$ resolution
difficult. When combining the methods, however, we note that $b(\xi)$
and $\bar b(\xi)$ are invariant under $\gamma \leftrightarrow
\delta_B$, whereas $a(\xi)$ and $\bar a(\xi)$ are invariant under
$\gamma \leftrightarrow \delta_B + \delta_D$. The $S_{\rm exchange}$
ambiguity is thus resolved in a single $B^+$ and $D$ decay mode in
which $\delta_D$ is far enough from 0 or $\pi$.

\section{Signal and Background Estimates}

We proceed to estimate the sensitivity of the $\gamma$ measurement
combining the ADS and GW methods, at a future, symmetric $e^+e^-$
$B$-factory, operating at the $\Upsilon$(4S) resonance. The detector
configuration is taken to be similar to that of
CLEO-III\cite{ref:cleo3}. The integrated luminosity is $600 \ {\rm
fb}^{-1}$, corresponding to three years of running at the full
luminosity of $3 \times 10^{34}$~cm$^{-2}$~s$^{-1}$\cite{ref:cesr4}
with an effective duty factor of 20\%.

Crucial to evaluating the measurement sensitivity is a reasonably
realistic estimate of the background rate in the measurement of $a_m$,
whose statistical error dominates the $\gamma$ measurement error,
$\Delta \gamma$. We estimated the background by applying
reconstruction criteria to Monte Carlo events generated using the
full, GEANT-based\cite{ref:geant} CLEO-II detector simulation. The
event sample consisted of about $19\times 10^6$ $e^+e^-\rightarrow
B\bar B$ events and $14\times 10^6$ continuum $e^+e^-\rightarrow q\bar
q$ events, where $q$ stands for a non-$b$ quark. Since the full
simulation did not include a silicon vertex detector or \u{C}erenkov
particle identification system, these systems were simulated using
simple Gaussian smearing. The \u{C}erenkov detector was taken to cover
the polar region $|\cos\theta| < 0.71$.

$D^0$ candidates (reference to the charge conjugate modes is implied)
were reconstructed in the final states $K^- \pi^+$, $K^- \pi^+ \pi^0$,
and $K^- \pi^+ \pi^- \pi^+$. The $\pi^0$ and $D^0$ candidate invariant
masses were required to be within $2.5$ standard deviations ($\sigma$)
of their nominal values. A Dalitz plot cut was applied in the $K^-
\pi^+ \pi^0$ mode to suppress combinatoric background.  The $B^+$
candidate energy was required to be within $2.5\sigma$ of the beam
energy. The beam-constrained mass, $\sqrt{E_b^2 - P_B^2}$, where $E_b$
is the beam energy and $P_B$ is the momentum of the $B^+$ candidate,
was required to be within $2.5\sigma$ of the nominal $B^+$ mass.
Since the $K^+$ and the $D^0$ fly back-to-back, all charged daughters
of the $B^+$ candidate were required to be consistent with originating
from the same vertex point.  Continuum background was suppressed by
applying cuts on the cosine of the angle between the the sphericity
axis of the $B^+$ candidate and that of the rest of the event, and on
the output of a Fischer discriminant~\cite{ref:b2dk}. In background
events, the reconstructed $K^+$ and $K^-$ come from two different
$D$~mesons, or are due to $s\bar s$ popping, while signal events often
contain a third kaon, originating from the other $B$ meson in the
event. As a result, 90\% of the background events are rejected by
requiring that an additional $K^-$ or $K_S$ be found in the event and
be inconsistent with originating from the $B^+$ candidate vertex.

With the above event selection criteria, we find that continuum events
account for over 80\% of the remaining background, with a rate of 7
events per $10^8$ charged $B$ mesons produced. This is comparable to
the expected signal yield. Under such low signal, high background
conditions, significant improvement is obtained by conducting a
multi-variable maximum likelihood fit. In this technique, cuts on the
continuous variables are greatly loosened, and the separation of
signal from background is achieved by use of a probability density
function, which describes the distribution of the data in these
variables. As has been the case in several CLEO analyses of rare $B$
decays, we assume that the effective background level in the
likelihood analysis, $B$, as inferred from the signal statistical
error, $\Delta S = \sqrt{S + B}$, will be similar to the level
obtained with the Monte Carlo simulation. Signal efficiency will
increase, however, due to the looser selection criteria.

The expected number of $B^+\rightarrow fK^+$ signal events is
%%%%%%%%
\begin{equation}
N_a = N_{B^+} \;
	a(\xi)\;
	\epsilon(K^+f),
	\label{eq:N_a}
\end{equation}
%%%%%%%%%%%%%%%%%%%
where $N_{B^+}$ is the number of $B^+$~mesons produced, and
$\epsilon(K^+ f)$ is the probability that the final state be detected
and pass the loosened selection criteria of the likelihood analysis.
For given values of $\delta_D$, $\delta_B$ and $\gamma$, we calculate
$a(\xi)$ using the $D^0\rightarrow K^-\pi^+,\ K^-\pi^+\pi^0,\
K^-\pi^+\pi^-\pi^+$ branching fractions from~\cite{ref:pdg},
Equation~(\ref{eq:dcs}), ${\cal B}(B^+\rightarrow \bar D^0 K^+) =
2.57\times 10 ^{-4}$~\cite{ref:b2dk}, and ${\cal B}(B^+\rightarrow
D^0K^+)= 2.3\times 10^{-6}$ (obtained from ${\cal B}(B^+\rightarrow
\bar D^0 K^+)$ and the values used in Equation~(\ref{eq:GW-trouble})).

To estimate the efficiency $\epsilon(K^+ f)$, we start with the values
in~\cite{ref:b2dk}, 44\% for the $K^-\pi^+$ mode, 17\% for the
$K^-\pi^+\pi^0$ mode, and 22\% for the $K^-\pi^+\pi^-\pi^+$ mode.
These are multiplied by the efficiency of finding the third kaon
(45\%), and the particle-ID efficiency (68\%). The particle-ID
efficiency is composed of the probability that a well-reconstructed
$K^+$ be in the particle-ID system's fiducial region (83\%), and that
half the $K^-$ daughters of the $D$ meson also be in the fiducial
region. The momentum of the other half allows good identification
using specific ionization, as does the momentum of the third kaon in
most events. An additional efficiency loss of 10\% is assumed due to
non-Gaussian tails, \u{C}erenkov ring overlaps, etc. The final
efficiencies are 13\% for the $K^-\pi^+$ mode, 5\% for the
$K^-\pi^+\pi^0$ mode, and 7\% for the $K^-\pi^+\pi^-\pi^+$ mode.

Since $b_m \gg a_m$, suppression and accurate knowledge of the
background in the measurement of $b_m$ is much less critical. Starting
from the continuum background level in~\cite{ref:b2dk} and applying
vertex and particle-ID criteria, we arrive at a rate of 60 background
events per $10^8$ charged $B$ mesons. The number of signal events
observed in this channel is
%%%%
\begin{equation}
N_b = N_{B^+} \;
	b(\xi)\;
	\epsilon(K^+) \; 
	\sum_i {\cal B}(D^0 \rightarrow c_i) \; \epsilon(c_i), 
\label{eq:N_b}
\end{equation}
%%%%%%%%%%%%%%%%%%%
where $\epsilon(K^+)$ is the efficiency for detecting the $K^+$ with
the particle-ID criteria described above, and $c_i$ are CP-eigenstate
decay products of $D_{1,2}$. Using Table~\ref{tab:cp-modes}, we obtain
$\sum_i {\cal B}(D^0 \rightarrow c_i) \; \epsilon(c_i) = 0.011$.
 
\section{Measurement Sensitivity}

To estimate the measurement sensitivity for given values of the
``true'' parameters $\xi = \xi^0$, we compute the average numbers of
observed signal events using Equations~(\ref{eq:N_a})
and~(\ref{eq:N_b}). 
An integrated luminosity of $600 \ {\rm fb}^{-1}$ yields 
$N_{B^+} = 640\times 10^6$. We assume
that statistics will effectively triple if, in addition to 
$B^+\rightarrow D^0 K^+$, one uses the modes
$B^+\rightarrow D^0 K^{*+}$, 
$B^+\rightarrow D^{*0} K^+$,
$B^+\rightarrow D^{*0} K^{*+}$,
$\bar B^0\rightarrow D^0 K^{*0}$
and 
$\bar B^0\rightarrow D^{*0} K^{*0}$.
We therefore take $N_{B^+} = 1900\times 10^6$. 
The resulting $N_a$, $N_b$ and their CP-conjugates determine the
experimental quantities $a_m$, $b_m$, $\bar a_m$ and $\bar b_m$ in the
average experiment, ie., the experiment in which statistical
fluctuations vanish. The minimization package MINUIT~\cite{ref:minuit}
is then used to find the parameters $\xi$, for which $\chi^2(\xi)$ is
minimal in this experiment. Since the measurement is expected to be
statistics-limited, only statistical errors are used to evaluate
$\chi^2(\xi)$. 

\begin{figure}[p]
\centering
\epsfxsize=4.5in
\epsffile{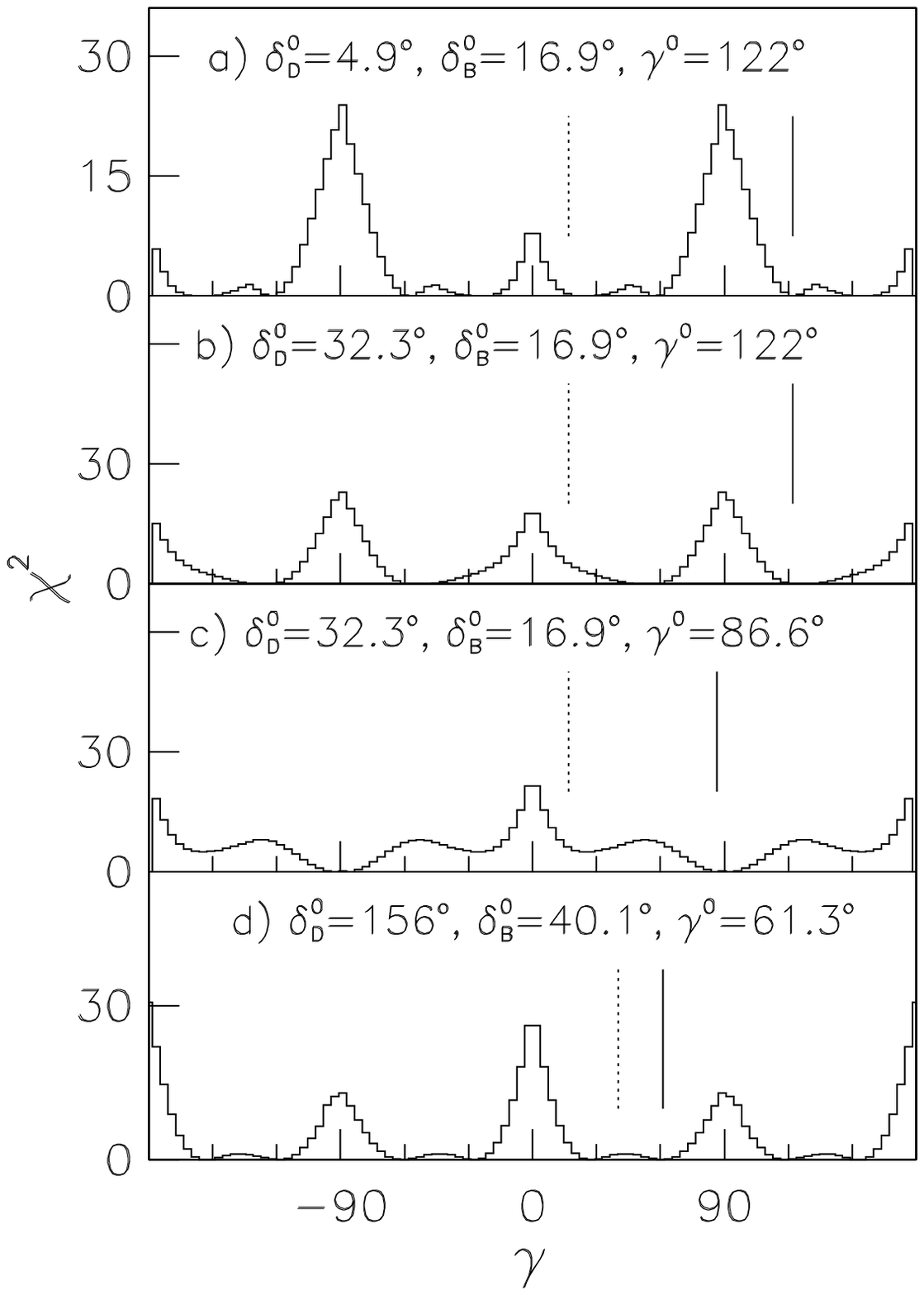}
\caption{$\chi^2(\xi)$ as a function of $\gamma$ for different values
of the actual phases, $\delta_D^0$, $\delta_B^0$, $\gamma^0$. For each
value of $\gamma$, $\chi^2(\xi)$ is minimized with respect to ${\cal
B}(B^+\rightarrow D^0 K^+)$, $\delta_D$ and $\delta_B$. The points
$\gamma = \gamma^0$ and $\gamma = \delta_B^0$ are shown by a solid and a
dotted line, respectively. Some asymmetry and noise are due to the
dependence of the fit on the initial $\xi$ values. {\bf a)} The 8-fold
ambiguity of Equations~(\ref{eq:ambig}) and~(\ref{eq:ambig2}) is
demonstrated for small $\delta_D^0$. 
%  The $S_{\rm exchange}$
%  ambiguity is only approximate here, since ${\cal B}(B^+\rightarrow D^0
%  K^+)$ varies with $\gamma$. 
{\bf b)} Increasing $\delta_D$, the
$S_{\rm exchange}$ ambiguity is resolved. {\bf c)} With $\gamma$ close
to $90^\circ$, the $S_\pi$ and $S_{\rm sign}$ ambiguities
overlap. {\bf d)} The $S_{\rm exchange}$ ambiguity is resolved, but an
accidental ambiguity shows up at $\gamma \approx 28^\circ$, with
${\cal B}(B^+\rightarrow D^0 K^+)$ at approximately $4/3$ its input
value.}
\label{fig:gamma-scans}
\end{figure}

To demonstrate ambiguities, the trial value of $\gamma$ is stepped
between $-180^\circ$ and $180^\circ$, and $\delta_{D}$, $\delta_{B}$
and ${\cal B}(B^+\rightarrow D^0 K^+)$ are varied by MINUIT so as to
minimize $\chi^2(\xi)$. Such $\gamma$ scans are shown in
Figure~\ref{fig:gamma-scans} for cases of particular interest. Evident
from these scans is the fact that a large $\partial^2 \chi^2(\xi) /
\partial \gamma^2$ at the input value $\gamma = \gamma^0$ does not
guarantee that $\chi^2(\xi)$ will obtain large values before dipping
into a nearby ambiguity point. As a result, the quantity that
meaningfully represents the measurement sensitivity is not
$\Delta\gamma$, but $f_{\rm exc}$, the fraction of $g$ which is
excluded by the $B\rightarrow DK$ measurement, ie., for which
$\chi^2(\xi) > 10$. The larger the value of $f_{\rm exc}$, the greater
the a-priori likelihood that predictions of $\gamma$ based on new
physics-sensitive experiments will be inconsistent with the
$B\rightarrow DK$ measurement, leading to the detection of new
physics.

To evaluate $f_{\rm exc}$, 540 Monte Carlo experiments were generated,
using randomly selected input values in the range $\gamma^0\in g$,
$-180^\circ < \delta_D^0 < 180^\circ$, $-180^\circ <\delta_B^0 <
180^\circ$ (Note that in reality, the CP-conserving phases will be
different in the different decay modes).  Depending on the input
phases, the numbers of observed signal events varied between $700
< N_b < 1050$, $0 < N_a < 130$. For each set of phases, a $\gamma$
scan was conducted in the range $\gamma \in g$, and $f_{\rm exc}$ was
taken to be the fraction of the area of the scan for which
$\chi^2(\xi) > 10$.

\begin{figure}[p]
\centering
\epsfxsize=3.3in
\epsffile{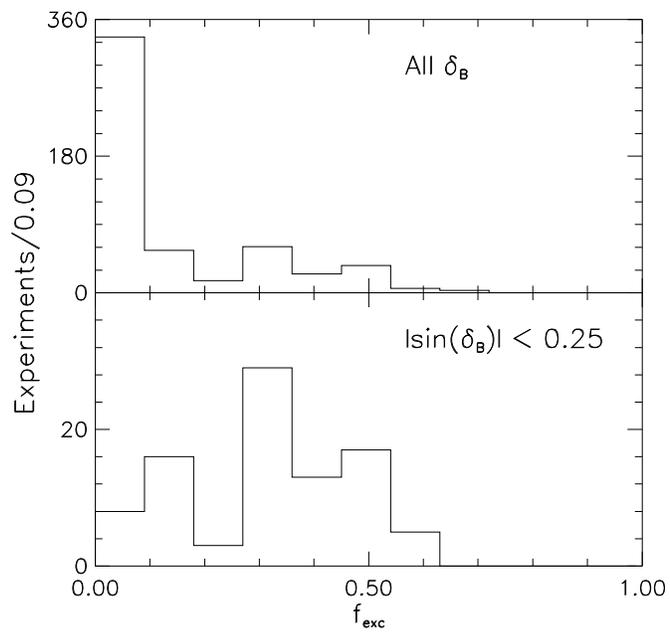}
\caption{The $f_{\rm exc}$ distribution of all Monte Carlo experiments
conducted, and experiments with $|\sin(\delta_B)| < 0.25$.}
\label{fig:f_exc}
\end{figure}

The $f_{\rm exc}$ distribution of the 540 random experiments is shown
in Figure~\ref{fig:f_exc}. Also shown is the distribution of the 91
experiments for which $|\sin(\delta_B)| < 0.25$. $f_{\rm exc}$ tends
to be larger in this case, since small values of $\chi^2(\xi)$
associated with the $S_{\rm exchange}$ ambiguity (even if the
ambiguity is resolved) are pushed away from the center of $g$. Since
the distributions of phases used in the Monte Carlo experiments cannot
be expected to represent the actual phases in nature, it is not
meaningful to study the $f_{\rm exc}$ distribution in
detail. Nevertheless, Figure~\ref{fig:f_exc} indicates that this
measurement may reduce the allowed region of $\gamma$ by as much as
70\%.

\section{Discussion and Conclusions}

We have studied in detail the measurement of $\gamma$ using
$B\rightarrow DK$ at a symmetric $B$~factory. Use of this measurement
to detect new physics effects is complicated by low statistics and an
ambiguity which is at least 8-fold, not 4-fold as often stated. We
show that combining the ADS and GW methods helps resolve the $S_{\rm
exchange}$ ambiguity and decreases the statistical error, compared
with the ADS method alone. The ambiguities associated with the $S_{\rm
sign}$ and $S_\pi$ symmetries are irremovable in measurements of this
kind. Even when the $S_{\rm exchange}$ ambiguity is in principle
resolved, in practice it still deteriorates the measurement by
reducing $\chi^2(\xi)$ (or other experimental quantity of significance).

Being ambiguity-dominated, the sensitivity of future experiments
should be evaluated in terms of the exclusion fraction $f_{\rm exc}$,
rather than the weak phase error $\Delta\gamma$. With a luminosity of
$600 \ {\rm fb}^{-1}$, we find that the $B\rightarrow DK$ measurement
can exclude up to about $f_{\rm exc} \lesssim 0.6$ of the currently-allowed
range of $\gamma$.

With $3\times 10^8$ $B$~mesons, 100\% efficiency and no background, JK
find $\Delta\gamma$ in their method to be between about $5^\circ$ and
$30^\circ$ for $40^\circ < \gamma < 100^\circ$. Using more realistic
estimates and noting out comments above, one would conclude that
combining their method with the ADS and GW methods, while probably
useful for the actual experiment, will not result in a dramatic change
in the predictions of our analysis.

\section{Acknowledgments}
I am grateful to my colleagues at the CLEO collaboration for
permitting the use of the excellent Monte Carlo sample which they have
worked hard to tune and produce; to David Asner and Jeff Gronberg for
sharing their knowledge of the performance of the CLEO silicon vertex
detector; and to Michael Gronau and Yuval Grossman for discussions and
useful suggestions. This work was supported by the U.S. Department of
Energy under contracts DE-AC03-76SF00515 and DE-FG03-93ER40788, and by
the National Science Foundation.

%\begin{figure}[p]
%\centering
%\epsfxsize=6in
%\epsffile{gw-triangle-3590198-002.ps}
%\caption{The triangle relation
%$D^0_+ \equiv {1\over\sqrt{2}} \left(D^0 + \bar D^0\right)$
%applied to the $B^+$ and $B^-$~decay amplitudes. Similar triangles
%exist for $D^0_- \equiv {1\over\sqrt{2}} \left(D^0 - \bar D^0\right)$.}
%\label{fig:GW-triangle}
%\end{figure}

% This figure is produced by running sim/submit-2.plb from 751/ to create the
% plots, and then running ../tex/plb2.mnf in mn_fit. ../tex/plb2.mn is used 
% by plb2.mnf.
%

% These figures are produced by running sim/submit-2-2.plb from 751/ 
% and then running ../tex/plb2-2.mnf in mn_fit.
% 
%
%\begin{figure}[p]
%\centering
%\epsfxsize=3.3in
%\epsffile{nevents.ps}
%\caption{The numbers of observed Monte Carlo experiments signal events 
%in the $N_a$ and $N_b$ channels, vs. their CP-conjugates.}
%\label{fig:nevents}
%\end{figure}

\begin{table}
\begin{center}
\begin{tabular}{lccc}
$t$ & ${\cal B}(D^0\rightarrow t)$ & $\epsilon(t)$ & 
	${\cal B} \times \epsilon$ \\
\hline
$K_S \; \pi^0$                               & 0.011   & 0.22  & 0.0024 \\
$K_S \; \eta(\rightarrow \gamma\gamma) $     & 0.0036  & 0.087 & 0.0003 \\
$K_S \; \rho^0 $                             & 0.0061  & 0.28  & 0.0017 \\
$K_S \; \omega(\rightarrow \pi^+\pi^-\pi^0)$ & 0.011   & 0.13  & 0.0014 \\
$K_S \; \eta' (\rightarrow \pi^+\pi^-\eta) $ & 0.0086  & 0.062 & 0.0005 \\
$K_S \; \eta' (\rightarrow \rho^0\gamma) $   & 0.0086  & 0.068 & 0.0006 \\
$K_S \; \phi(\rightarrow K^+K^- )$           & 0.0043  & 0.14  & 0.0006 \\
$K^+  K^- $                                  & 0.0043  & 0.64  & 0.0028 \\
$\pi^+  \pi^- $                              & 0.0015  & 0.64  & 0.0010 \\
\hline
total                                        &          &      & 0.011  \\
\end{tabular}
\end{center}
\caption{Branching fractions~\protect\cite{ref:pdg} of $D^0$ decays to
CP-eigenstates, assumed reconstruction efficiencies, and their
products. Efficiencies include sub-mode branching fractions, such as
$K_S \rightarrow \pi^+\pi^-$, and are constructed assuming 80\% track
efficiency and 50\% $\pi^0$ efficiency.}
\label{tab:cp-modes}
\end{table}

\end{document}